\documentclass[runningheads]{llncs}
\usepackage[T1]{fontenc}
\usepackage{graphicx}
\usepackage{url}

\usepackage{amsmath}
\usepackage{booktabs}
\usepackage{array}
\usepackage{placeins}
\usepackage[most]{tcolorbox}
\definecolor{terra}{HTML}{DC7A52}
\definecolor{navy}{HTML}{38425C}

\newcommand{\plotwidth}{\textwidth}

\begin{document}

\title{Determinants and Limits of LLM Security-Tool Orchestration: A Study with HexStrike-AI}
\titlerunning{Determinants and Limits of LLM Security-Tool Orchestration}

\author{Romain Gerard\inst{1}\thanks{R. Gerard and A. Zeghaider contributed equally to this work.} \and Assmaa Zeghaider\inst{1} \and Yan Guo\inst{2}}
\authorrunning{R. Gerard et al.}
\institute{University of Science and Technology of China, Hefei, China \and
Suzhou Institute for Advanced Research, University of Science and Technology of China\\
\email{guoyan@ustc.edu.cn}}

\maketitle

\begin{abstract}
Large language model agents driving security tool suites over the Model Context Protocol are increasingly common. Yet the factors that bound their capability remain poorly characterized: how much depends on the model versus the client that drives it, whether constraining the agent to the orchestrator's own tools helps, and where capability is limited by reasoning rather than by missing tools. Using HexStrike-AI, an open-source orchestrator that exposes 150+ tools, as a testbed, we follow a methodology that evaluates the system, diagnoses its failures, and applies targeted improvements. We run 86 picoCTF challenges across seven categories and three difficulty tiers, under three tool-access regimes and three model/client configurations (774 trials). We then apply corrections to existing tools, agent-behavior changes, and eleven new capability tools, and re-run the previously-unsuccessful trials. The diagnosis isolates the driving client as a first-order factor for a fixed model (a 2.1$\times$ gap between two DeepSeek clients) and a monotonic difficulty gradient, with the largest gains in the mid tier. The overall solve rate rises from 55.4\% to 72.0\%, and every configuration improves significantly (paired McNemar $p < 0.001$, non-overlapping 95\% confidence intervals). The residual failures are reasoning- or environment-bound rather than missing-tool. A 60-run stability sub-study finds single-run verdicts reproducible (17/20 unanimous). We discuss what the results imply for how such orchestrators should be evaluated, and we are explicit about the limits: the study uses a single benchmark, the fixes were tuned on the same challenges they were evaluated on, and the client effect is demonstrated for one model only, so its generality to other models remains a hypothesis.

\end{abstract}

\section{Introduction}
\begin{figure}[t]
    \centering
    \includegraphics[width=\plotwidth]{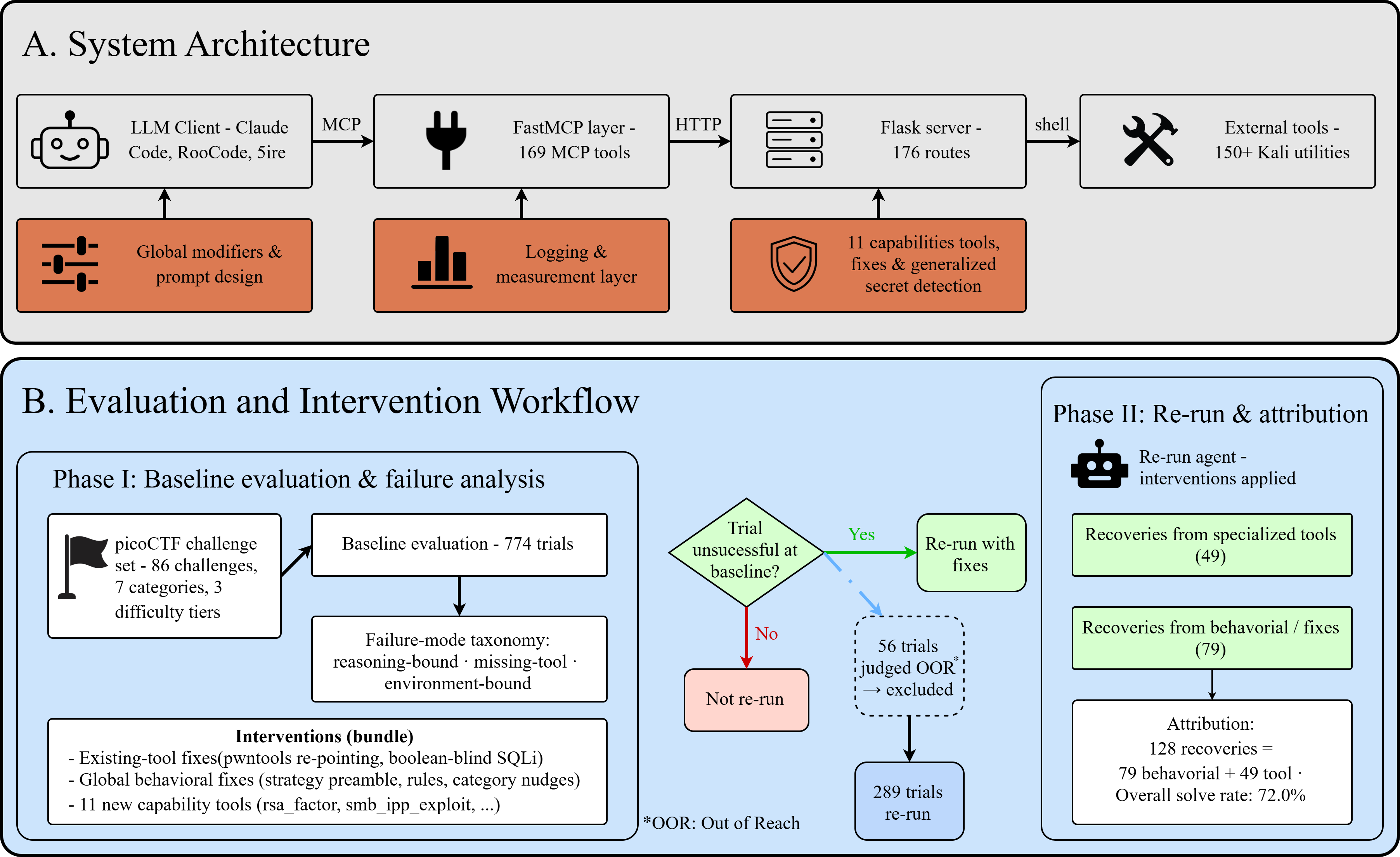}
    \caption{HexStrike-AI architecture and our evaluation-and-intervention workflow. (A) The upstream pipeline routes an LLM client through a FastMCP layer (169 MCP tools) and a Flask server (176 routes) out to 150+ external Kali utilities, over MCP, HTTP, and shell; our contributions (shaded) add the global modifiers and prompt design on the client, the logging and measurement layer, and the eleven capability tools, fixes, and generalized secret detection on the server. (B) The study workflow: a baseline over 774 trials with a failure-mode characterization (Phase I), then the bundled intervention applied on re-run of only the previously-unsuccessful trials (56 judged out of reach are excluded and stay counted as failures), leaving 289; each recovery is attributed to a specialized tool (49) or to behavioral and default fixes (79), for 128 recoveries and a 72.0\% overall solve rate (Phase II).}
    \label{fig:fig7}
\end{figure}

Large language models are increasingly deployed as agents that plan and call external tools~\cite{toolformer,react}. Offensive security is a natural application: penetration testing and capture-the-flag (CTF) solving are tool-heavy, multi-step tasks~\cite{pentestgpt,nyuctf} in which an agent must choose, invoke, and chain specialized utilities. A growing class of orchestrators exposes large tool suites to an LLM over the Model Context Protocol (MCP)~\cite{mcp}, letting one agent drive scanners, exploit frameworks, and forensics utilities. HexStrike-AI (hereafter HexStrike) is one such open-source orchestrator: it wraps 150+ security tools and exposes roughly 170 of them to an LLM client through a Flask backend and a FastMCP interface (Figure~\ref{fig:fig7}). Systems of this kind are multiplying, yet how well they work, and what determines the performance, is not established.

The open questions are concrete: how much of an agent's success is set by the model versus the client that drives it, whether constraining the agent to the orchestrator's own tools helps, and where capability runs out. The last question also asks whether the residual failures are due to missing-tool or reasoning. Two further dimensions intersect with this analysis: the reproducibility of single-run verdicts for drawing stable conclusions, and the generalizability of a CTF-centric orchestrator to real-world target environments. We frame the central three as research questions:
\begin{itemize}
    \item \textbf{RQ1.} When an LLM agent drives a security-tool orchestrator, what governs its success: the underlying model, the client (harness) that drives it, or the tool-access regime, and which dominates?
    \item \textbf{RQ2.} When such a system is improved, does recovery come more from adding specialized tools or from correcting the agent's default behavior, and where do the gains concentrate?
    \item \textbf{RQ3.} Where does such an orchestrator's capability run out, and are the residual failures bounded by missing tools or by reasoning and environment?
\end{itemize}

We study these questions empirically, treating HexStrike-AI as a testbed rather than a system to introduce; a recent cross-framework evaluation found it the strongest performer among five open-source penetration-testing frameworks~\cite{decoupled-pentest}, supporting its choice as a representative subject. We evaluate it across three model/client configurations and three tool-access regimes on 86 picoCTF challenges (774 trials), then characterize its failures, apply targeted improvements, and re-run the previously-unsuccessful trials to measure the effect.

This paper makes the following contributions:
\begin{enumerate}
    \item An evaluation methodology for an LLM security-tool orchestrator (challenge set, three tool-access regimes, multi-model/client design, scoring, and logging) and the resulting 774-trial dataset.
    \item A characterization of an orchestrator's failure modes: a difficulty-and-category map of where capability runs out and a taxonomy of why the residual failures resist tooling.
    \item An intervention-with-attribution study that separates tooling from behavior: corrections, agent-behavior changes, and eleven new capability tools recover 128 trials, 79 from behavioral and default fixes and 49 from specialized tools.
    \item A measurement of the driving client as a first-order factor for a fixed model: holding the DeepSeek model constant, two clients differ by 2.1$\times$, so a model's solve rate is meaningful only when the client that produced it is named.
\end{enumerate}

\section{Related Work}
A growing body of work drives LLM agents through penetration-testing workflows, beginning with PentestGPT~\cite{pentestgpt} and extending to multi-agent and benchmark-driven systems~\cite{vms,autopt-isozaki}. Several are evaluate-and-improve studies that characterize failures and add targeted augmentations, mirroring our intervention design. Most directly, PentestGPT~v2~\cite{pentestgpt-v2} sorts agent failures into two kinds: capability gaps that engineering can remove, and planning or state-management barriers that persist regardless of tooling or model. This independently parallels our finding: our recoveries came from tools and default-behavior fixes, whereas the hard failures that remain are reasoning- or environment-bound. Methodological work~\cite{bench-practices} calls for baselines and richer metrics and notes that CTF challenges may not represent real engagements, the caveat we also raise. Reliability, by contrast, is rarely measured: a 400-run study with target, prompt, and orchestrator fixed shows offensive-LLM behavior varies by model and by run~\cite{attacker-consistency}, motivating our stability check. We study an existing orchestrator rather than proposing a new agent, and isolate the driving client from the model.

CTF challenges are the dominant proxy for offensive capability, supported by open benchmarks such as NYU CTF Bench~\cite{nyuctf} and Cybench~\cite{cybench} and a growing line of CTF-solving agents~\cite{craken,ctftiny}. Reported solve rates vary widely with model and method, matching the spread we observe across client configurations (27.9--89.9\%). Closest to our thesis, EnIGMA~\cite{enigma} shows that adding interactive tools and interfaces substantially raises an agent's CTF solve rate, the same mechanism behind our tool corrections and capability tools. Other work moves beyond binary flag capture toward partial-credit scoring~\cite{deepred,ctftiny}, paralleling our PARTIAL verdict, and examines sensitivity to sampling and hyperparameters~\cite{vms,ctftiny}, which our stability sub-study addresses. A negative result on an MCP-grounded CTF agent~\cite{skills-negative} finds that, once the tool layer returns strict schema-validated feedback, added procedural guidance helps little, echoing our finding that the gains came from the tool layer itself rather than from client-side sophistication.

LLM tool use rests on foundational work, from reasoning-and-acting loops~\cite{react} and self-taught tool use~\cite{toolformer} to function calling over large API collections~\cite{gorilla}; HexStrike exposes its tools over the Model Context Protocol~\cite{mcp}. A more recent line argues that the execution harness around the model, which SWE-agent terms the agent-computer interface~\cite{swe-agent}, is itself a first-order determinant of performance. With the model fixed, the harness can shift results as much as the model and even reverse rankings~\cite{harness-bench,harness-disclosure}; the security-domain instance makes the same point on CTF challenges~\cite{csi-harness}. Related work measures behavioral reproducibility across repeated tool-calling runs~\cite{agent-consistency}, with agents selecting the same tools but varying their arguments across runs. Our client-as-a-first-order-factor result is a concrete security-domain measurement in this space: holding the DeepSeek model fixed, two MCP-capable clients differ by 2.1$\times$.

\section{Method}

Our design follows from the research questions. For RQ1, we cross three model/client configurations with three tool-access regimes, with the DeepSeek model fixed across two clients to isolate the client effect. For RQ2, we run a baseline, a bundled intervention, and a re-run of only the previously-unsuccessful trials, attributing each recovery to general behavior or to a specific tool. For RQ3, we build a category-by-difficulty success map and characterize the residual failures. Inferential tests and a stability sub-study bound how far the conclusions can be trusted.

\subsection{Challenge Set}
We use 86 picoCTF challenges across seven categories (Web, Cryptography, Binary Exploitation, Forensics, Reverse Engineering, General Skills, Blockchain) and three difficulty tiers (27 Easy, 31 Medium, 28 Hard). Challenges were used as published, with live instances provisioned through picoCTF's panel.

\subsection{Experiment Design}
Each challenge was attempted by a given model/client under three regimes:
\begin{itemize}
    \item \textbf{Exp 1 (Free solve).} Any tool, native or HexStrike; measures the agent's standalone capability.
    \item \textbf{Exp 2 (HexStrike-ranked).} Restricted to HexStrike tools (native tools forbidden), with a ranked preference ordering among them.
    \item \textbf{Exp 3 (HexStrike-strict).} Restricted to HexStrike tools under a stricter prohibition than Exp 2. The baseline prompt forbids the agent's own tools and logic, and the intervention later hardens this wording further, though native tools are never technically disabled (Section~\ref{sec:adherence}).
\end{itemize}

Exp 1 isolates model capability; Exp 2/3 isolate how far HexStrike's own tooling carries the agent. Both constrained regimes prohibit native tools by instruction rather than by technically disabling them, so how well the prohibition held is itself a measured outcome (Section~\ref{sec:adherence}). Each run used a fixed two-step prompt: a setup step (model/client identity, challenge metadata, timer start) and a task step. At baseline the task step was deliberately minimal, a one-line role instruction that varied only by regime followed by the challenge name and description. The hypothesis-first strategy preamble, the always-on rules, and the per-category tool nudges were added later as part of the intervention (Figure~\ref{fig:prompt}, Section~\ref{sec:fixes}), not at baseline; the figure shows both the baseline and the post-fix task prompt. All prompts are produced by a prompt generator and the verbatim text is in the repository. The design uses three model/client configurations: Claude (Sonnet 4.6) via Claude Code, and DeepSeek (\texttt{deepseek-chat}) via RooCode and via 5ire. Holding the model fixed across the two DeepSeek clients isolates the client effect, and the full matrix of $86 \times 3$ regimes $\times 3$ configurations yields 774 trials.

\begin{figure}[t]
    \centering
    \tcbset{promptbox/.style={
        boxrule=0.5pt, coltitle=white, fonttitle=\bfseries\footnotesize,
        left=5pt, right=5pt, top=2pt, bottom=2pt, before skip=3pt, after skip=4pt}}
    \begin{minipage}{0.9\linewidth}
    \begin{tcolorbox}[promptbox, colframe=black!45, colback=black!3,
        colbacktitle=black!55,
        title={Step 1, setup\hfill\mdseries\itshape both phases}]
    \scriptsize
\begin{verbatim}
set_llm_identity(model, client)
set_ctf_metadata(name, difficulty, type)
start_timer()
\end{verbatim}
    \end{tcolorbox}
    \begin{tcolorbox}[promptbox, colframe=terra, colback=terra!8,
        colbacktitle=terra!80!black,
        title={Step 2, task prompt (minimal)\hfill\mdseries\itshape baseline}]
    \scriptsize
\begin{verbatim}
[Exp 1] You are a cybersecurity engineer. Solve this CTF
  using proven engineering methods.
[Exp 2] ...who ONLY uses Hexstrike tools; find the flag
  using hexstrike tools and ranking.
[Exp 3] ...Never use your own tools or your own logic;
  MUST follow hexstrike's logic.
(then: exercise name + description)
\end{verbatim}
    \end{tcolorbox}
    \begin{tcolorbox}[promptbox, colframe=navy, colback=navy!7,
        colbacktitle=navy,
        title={Step 2, added by the intervention\hfill\mdseries\itshape post-fix}]
    \scriptsize
\begin{verbatim}
MANDATORY FIRST STEP, before any tool, state:
  (a) the likely vulnerability class
  (b) your planned approach in 2-3 sentences
  (c) which tool you will try first and why
RULES (every run):
  - summarize each tool's output in 3 lines max
  - if a tool fails, diagnose WHY before switching
  - do NOT call web_request / source_code_read
    (they do not exist and always fail)
[+ if HARD/MEDIUM] call decompose_challenge(...) for a plan
[+ if HARD] after 3 tool calls, STOP and re-evaluate
\end{verbatim}
    \end{tcolorbox}
    \end{minipage}
    \caption{The two-step run prompt. Step 1 is a fixed setup (identity, metadata, timer), identical every run and in both phases. Step 2 is the task step: at baseline a one-line role instruction that varies only by regime, plus the challenge text; the intervention later adds the hypothesis-first block (a--c), the always-on rules, and a phased-decomposition call (the per-category tool nudges and the strengthened Exp 3 wording are omitted here for space). The verbatim text is produced by the prompt generator in the repository.}
    \label{fig:prompt}
\end{figure}

\subsection{Scoring and Measurement}
Each trial received a manual verdict (SUCCESSFUL, PARTIAL, or FAILED) recorded as a \texttt{TEST RESULT} line. SUCCESS required the correct flag, verified on picoCTF; PARTIAL denotes meaningful progress short of the flag, in the spirit of partial-credit CTF scoring~\cite{deepred,ctftiny}; FAILED otherwise. The reported success rate counts SUCCESSFUL only (PARTIAL is treated as non-success). Verdicts were assigned by the authors from the session logs. The only subjective boundary is PARTIAL versus FAILED, and because PARTIAL is folded into non-success throughout, that boundary does not affect any reported success rate.

A logging layer recorded, per session: the decision-engine tool chain, individual HexStrike tool calls, native (non-HexStrike) calls, a wall-clock timer, and the verdict. Native-tool calls under Exp 2/3 were logged to quantify leakage from the tool-access constraint.

\subsection{Intervention and Re-Run Protocol}
After establishing the baseline we applied targeted fixes: corrections and redirection of existing tools, global behavioral modifiers applied to every run, and eleven new capability tools. We then re-ran the previously-unsuccessful trials (those that had FAILED or PARTIALed) under the same model and client, but with the intervention applied, which includes the revised prompt (Section~\ref{sec:fixes}). From this re-run we excluded 56 trials judged out of reach, that is, challenges that would need a substantial new tool or external infrastructure, or infeasible effort, to move, leaving 289 re-run. The 56 excluded trials remain counted as failures in the post-fix totals, so removing them from the re-run set does not inflate the post-fix rate. Because only non-successes were re-run, any new SUCCESS is a real change from a recorded non-success; previously-successful trials were not re-run (this makes the paired before/after one-directional; see Section~\ref{sec:threats}).

\subsection{Fixes and Capability Tools}
\label{sec:fixes}
The intervention has three parts. First, \textbf{existing-tool and behavioral fixes} correct or redirect tools that already shipped. The most consequential redirects binary work toward the existing \texttt{pwntools} tool. The others correct the Python-execution, HTTP, and port-scan tools, add a boolean-blind SQLi extractor, and remove two advertised-but-nonexistent tools (\texttt{web\_request}, \texttt{source\_code\_read}). Second, four \textbf{global modifiers} replace the minimal baseline prompt and apply to every run: a hypothesis-first preamble, phased decomposition for harder challenges, a per-call confidence and next-tool signal, and a HexStrike-only constraint via the 5ire proxy (Figure~\ref{fig:prompt}). Third, \textbf{eleven new capability tools} each wrap an external capability for a category where the baseline most often stalled; Table~\ref{tab:t7} pairs each with the capability it adds and the challenge that motivated it.

\begin{table}[tb]
    \caption{Catalog of the Eleven New Capability Tools}
    \label{tab:t7}
    \centering
    \footnotesize
    \renewcommand{\arraystretch}{1.45}
    \setlength{\tabcolsep}{4pt}
    \begin{tabular}{>{\raggedright\arraybackslash}p{3cm}>{\raggedright\arraybackslash}p{5.4cm}>{\raggedright\arraybackslash}p{2.7cm}}
        \toprule
        Tool & Capability & Target Challenge(s) \\
        \midrule
        \texttt{rsa\_factor} & Recovers keys from weak or smooth RSA moduli & Crypto: Very Smooth \\
        \texttt{compression\_oracle} & CRIME/BREACH byte-by-byte recovery harness & Crypto: Compress and Attack \\
        \texttt{timing\_oracle} & Recovers a secret via response-time or instruction-count side channels & Forensics: SideChannel \\
        \texttt{sqli\_order\_oracle} & Boolean-blind SQL extraction (ORDER BY / CASE WHEN) & Web: ORDER ORDER \\
        \texttt{evtx\_parser} & Parses Windows .evtx logs for notable entries & Forensics: Event-Viewing \\
        \texttt{smb\_ipp\_exploit} & Reads files for secrets from SMB shares and IPP/CUPS printers & General: Printer Shares 2 and 3 \\
        \texttt{blockchain\_exploit} & Drives Foundry \texttt{cast} for access-control and reentrancy & Blockchain: Access\_Control, Smart\_Overflow, Reentrance \\
        ROP-chain builder & Builds a ROP chain (ROPgadget) and pwntools template & Binary/Hard \\
        Disk-image mount & Parses a disk image with The Sleuth Kit, listing allocated and deleted files & Forensics: DISKO 3, UnforgottenBits \\
        Encrypted-PCAP decryptor & Decrypts a captured session with tshark and a supplied key & Forensics: WebNet0, WebNet1 \\
        Headless XSS/CSRF chainer & Drives a headless browser to land an XSS/CSRF payload & Web: noted, secure-email-service \\
        \bottomrule
    \end{tabular}
\end{table}

All tools, old and new, share the same wiring (a Flask endpoint and a matching MCP wrapper, with shell arguments escaped and a fallback when a binary is absent). Several passed static checks yet failed only when run on the VM, so each tool was verified by execution rather than by compilation alone.

\subsection{Statistical Analysis}
For each configuration, we report the success rate $\hat{p} = x/n$ along with its 95\% Wilson score interval, which is known to maintain well-calibrated coverage even for the extreme success rates observed in our experiments:
\begin{equation}
    \frac{\hat{p} + \dfrac{z^2}{2n} \;\pm\; z\sqrt{\dfrac{\hat{p}(1-\hat{p})}{n} + \dfrac{z^2}{4n^2}}}{1 +
    \dfrac{z^2}{n}}, \qquad z = 1.96.
    \label{eq:wilson}
\end{equation}

We test each configuration's before-and-after improvement with McNemar's test for paired binary outcomes. Writing $b$ for trials that passed at baseline but failed afterward and $c$ for those that failed at baseline but passed afterward,
\begin{equation}
    \chi^2 = \frac{(b-c)^2}{b+c} \quad (1\ \text{df}).
    \label{eq:mcnemar}
\end{equation}
Because only previously-unsuccessful trials were re-run, no passing trial could regress, so $b = 0$ by construction; the statistic then reduces to an exact binomial sign test on the $c$ one-directional discordant pairs, with two-sided $p = 2(1/2)^{c}$.

We test the client effect (RooCode vs. 5ire, same model) with a two-proportion $z$-test on the baseline rates $\hat{p}_1$ and $\hat{p}_2$, using the pooled estimate $\hat{p} = (x_1 + x_2)/(n_1 + n_2)$:
\begin{equation}
    z = \frac{\hat{p}_1 - \hat{p}_2}{\sqrt{\hat{p}(1-\hat{p})\left(\dfrac{1}{n_1} +
    \dfrac{1}{n_2}\right)}}.
    \label{eq:twoprop}
\end{equation}

\subsection{Stability Sub-Study}
To characterize run-to-run variance~\cite{attacker-consistency,agent-consistency}, ten challenges (spanning all categories and difficulties, with a pass anchor and a fail anchor) were re-run three times on each DeepSeek client under Exp 1, for 60 runs in total. Sampling used the same client defaults as the main study, deliberately not pinned, so the measured variance reflects the study's own conditions. We report per-configuration agreement (identical verdict across the three runs) as the dataset's run-to-run noise floor.

\section{Results}

\subsection{Baseline}

Across the 774 baseline trials the three configurations were widely separated, from 78.7\% (Claude) down to 27.9\% (5ire), for 55.4\% overall (Table~\ref{tab:t1}). With the model held fixed, the client alone produced a large gap: the same DeepSeek model reached 59.7\% on RooCode but only 27.9\% on 5ire.

By tool-access regime (Table~\ref{tab:t2}), Claude performed best under the strict regime, RooCode was nearly flat, and 5ire was worst under the ranked regime, as though it treated ``ranked'' as optional.

\begin{table}[tb]
\centering
\begin{minipage}[t]{0.42\linewidth}
    \centering
    \caption{Baseline Outcomes by Configuration}
    \label{tab:t1}
    \footnotesize
    \setlength{\tabcolsep}{4pt}
    \begin{tabular}{lcccc}
        \toprule
        Configuration & S & F & P & Rate \\
        \midrule
        Claude/Sonnet 4.6 & 203 & 39  & 16 & 78.7\% \\
        DeepSeek/RooCode  & 154 & 95  & 9  & 59.7\% \\
        DeepSeek/5ire     & 72  & 172 & 14 & 27.9\% \\
        \midrule
        Overall           & 429 & 306 & 39 & 55.4\% \\
        \bottomrule
    \end{tabular}
\end{minipage}\hfill
\begin{minipage}[t]{0.42\linewidth}
    \centering
    \caption{Baseline Success Rate by Configuration and Experiment Regime}
    \label{tab:t2}
    \footnotesize
    \setlength{\tabcolsep}{4pt}
    \begin{tabular}{lccc}
        \toprule
        Configuration & Exp 1 & Exp 2 & Exp 3 \\
        \midrule
        Claude/Sonnet 4.6 & 76.7\% & 76.7\% & 82.6\% \\
        DeepSeek/RooCode  & 59.3\% & 59.3\% & 60.5\% \\
        DeepSeek/5ire     & 30.2\% & 19.8\% & 33.7\% \\
        \bottomrule
    \end{tabular}
\end{minipage}
\end{table}

\begin{figure}[tb]
    \centering
    \includegraphics[width=\plotwidth]{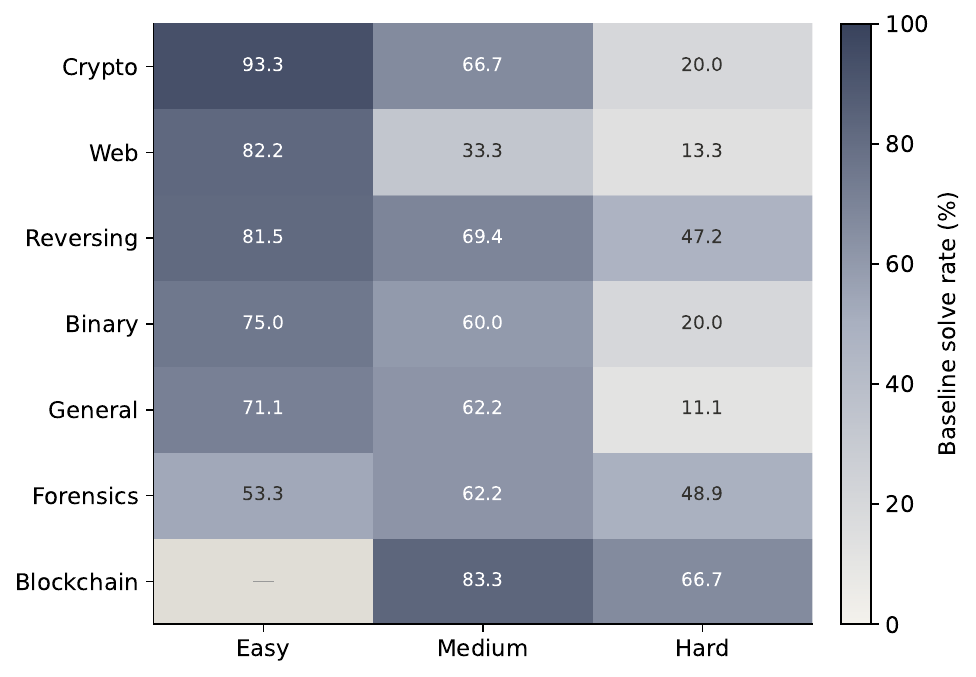}
    \caption{Baseline solve rate (\%) across the seven categories and three difficulty tiers. Capability falls off sharply with difficulty in almost every category, with the weakest cells in the Hard column where the baseline tooling is largely absent.}
    \label{fig:fig3}
\end{figure}

Capability fell off sharply with difficulty, from 75.7\% on Easy to 30.6\% on Hard, and the decline held within almost every category (Figure~\ref{fig:fig3}). The weakest category overall was Web at 43.0\%; the hardest cells were General/Hard (11.1\%) and Web/Hard (13.3\%), with Crypto/Hard and Binary/Hard at 20.0\%. Nine challenges failed in every configuration and every regime (a universal-failure set spanning multi-step web chains, disk forensics, and harder cryptography), marking where the baseline tooling was simply absent. Two genuine defects also surfaced. First, two advertised tools (\texttt{web\_request}, \texttt{source\_code\_read}) did not exist, so every call failed. Second, under the constrained regimes the models repeatedly reached for native tools despite the constraint, heaviest on 5ire and RooCode (quantified in Section~\ref{sec:adherence}).

\subsection{Effect of the Targeted Fixes}

\begin{figure}[tb]
    \centering
    \includegraphics[width=\plotwidth]{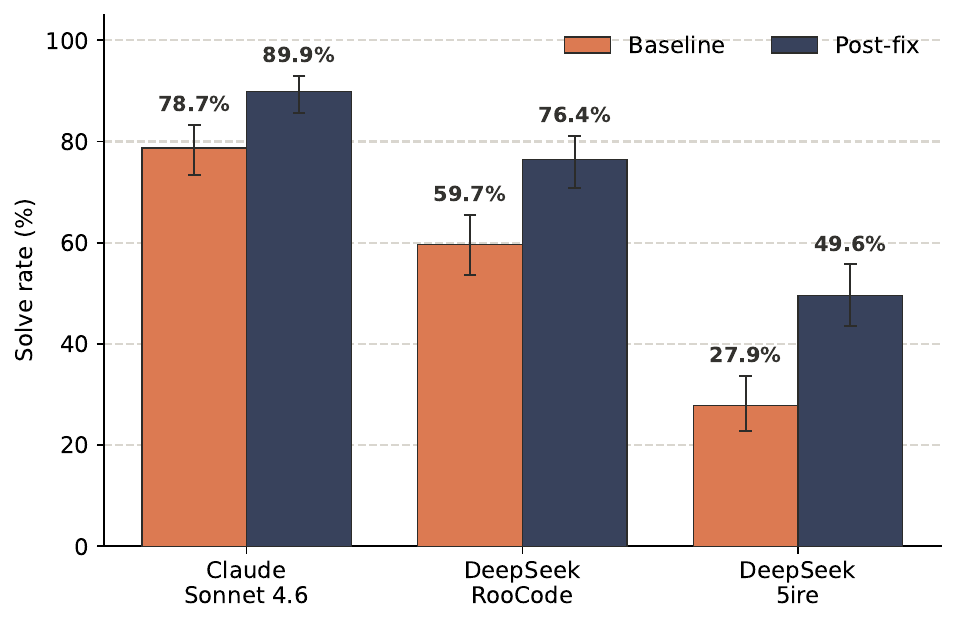}
    \caption{Solve rate before and after the fixes for each configuration, with 95\% Wilson confidence intervals. All three improve, and each configuration's before and after intervals do not overlap: Claude 78.7\% to 89.9\%, RooCode 59.7\% to 76.4\%, and 5ire 27.9\% to 49.6\%.}
    \label{fig:fig1}
\end{figure}

After the fixes, every configuration improved on every regime, and overall solve rate rose from 55.4\% to 72.0\% (557/774), with the largest gains on the weaker configurations (Claude +11.2pp, RooCode +16.7pp, 5ire +21.7pp; Figure~\ref{fig:fig1}). 5ire improved most under the strict regime (+31.4pp) and is the only configuration with a strong regime effect; Claude and RooCode lift uniformly across regimes (the cause is examined in Section~\ref{sec:adherence}).

In total, 128 previously-unsuccessful trials now pass. Of these, 79 came from the general behavioral and default fixes, and 49 used a specialized tool. The tools we built account for 24 of the 49: 22 from the eleven new capability tools, plus 2 from an added boolean-blind SQLi extractor. The other 25 came from redirecting the agent toward the existing \texttt{pwntools} tool, through a description rewrite plus a one-line prompt nudge rather than new code. This redirection was the single largest lever, and on its own it drove more recoveries than all the tools we built combined. The new tools were each invoked on the challenges they targeted; the lone exception, \texttt{timing\_oracle}, drew no calls on a signal-bound challenge rather than failing to invoke. Of the nine universal-failure challenges, four now solve in at least one configuration (including Printer Shares 2 and Credential Stuffing, solved by Claude) and three more reached partial progress, leaving only SRA and Secure Dot Product with no progress in any configuration.

\subsection{Failure Analysis}
\begin{figure}[tb]
    \centering
    \includegraphics[width=\plotwidth]{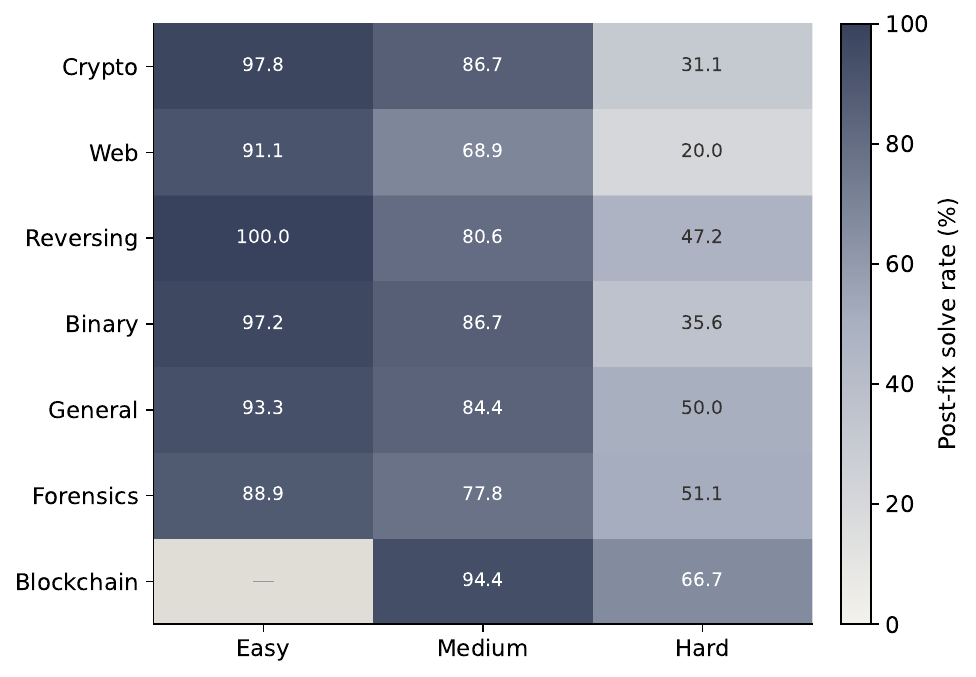}
    \caption{Post-fix solve rate (\%) across the seven categories and three difficulty tiers, the after to the baseline map in Figure~\ref{fig:fig3}. Every cell rises; the Hard column stays the floor while Easy and Medium approach saturation.}
    \label{fig:fig6}
\end{figure}

The post-fix solve rate remains monotonic in difficulty: Easy 94.2\%, Medium 81.7\%, Hard 39.7\%. Figure~\ref{fig:fig6} maps the post-fix rates across categories and difficulties, the after to the baseline map of Figure~\ref{fig:fig3}. The fixes lifted the middle most (Medium +21.5pp) and Hard least (+9.1pp). Recovery (the share of re-run trials that now pass) scales with configuration strength: among the re-run Hard trials, about 8\%, 20\%, and 38\% for 5ire, RooCode, and Claude respectively. The strongest configuration even cleared two Hard challenges with no dedicated capability tool, through general tools and reasoning. The residual failures fall into recognizable kinds: reasoning-bound (insight no tool supplies), interactive/real-time (a live race or multi-round protocol), multi-step stateful chains (every step must land), and one-off toolchains. What remains is reasoning- or environment-bound rather than a missing tool. SRA and Secure Dot Product, both Cryptography/Hard, are the genuine floor.

\subsection{Statistical Confidence}
For every configuration the baseline and post-fix 95\% Wilson intervals do not overlap, so the improvements are not interval artifacts (Figure~\ref{fig:fig1}). A paired McNemar test confirms each configuration improves at $p < 0.001$. Because only previously-unsuccessful trials were re-run, the test is one-directional by design: it establishes that the gains are not chance, not the absence of regressions among trials that already passed. The client effect is statistically significant: for DeepSeek the RooCode vs. 5ire gap is 2.1$\times$ at baseline (two-proportion $z$-test, $p \approx 3\times10^{-13}$), narrowing to about 1.5$\times$ after the fixes without closing. With only one model run on two clients, this supports a strong client effect for DeepSeek; the general claim that the client matters as much as the model is left as a hypothesis.

\subsection{Run-to-Run Reliability}
The stability sub-study (Table~\ref{tab:t6}) found RooCode unanimous on all ten challenges and 5ire unanimous on seven. The three that varied were all borderline Medium challenges, and each split two-to-one with a clear majority (RooCode solves those same three reliably). No challenge ever produced three different verdicts. Reproducibility scales with client strength, and the only variation sits on borderline-difficulty challenges for the weakest client; single-run verdicts elsewhere are stable.

\begin{table}[htb]
    \caption{Stability Sub-Study: Per-Configuration Agreement Across Three Runs}
    \label{tab:t6}
    \centering
    \begin{tabular}{lcc}
        \toprule
        Configuration & Unanimous (of 10) & Agreement \\
        \midrule
        DeepSeek/RooCode & 10 & 100\% \\
        DeepSeek/5ire    & 7  & 70\% \\
        \midrule
        Overall          & 17/20 & 85\% \\
        \bottomrule
    \end{tabular}
\end{table}

\subsection{Tool-Access Constraint Adherence}
\label{sec:adherence}
The constrained regimes (Exp 2 and Exp 3) forbid native tools by instruction, not by disabling them. The logs show the instruction was honored only loosely: the agents still issued 2,908 native \texttt{execute\_command} and 717 \texttt{Bash} calls, concentrated in the weaker 5ire and RooCode. A ``HexStrike-only'' figure is thus a soft constraint, not a clean partition, so constrained-regime results must be read alongside the leakage, not on their own. The leakage also explains the regime effect. 5ire gains most under the strict regime (+31.4pp, 33.7\% to 65.1\%) precisely because the stricter wording most curbs its native-tool fallback, pushing it onto the improved HexStrike tools it otherwise bypasses. Constraint adherence is itself a configuration-dependent confound that any evaluation of a constrained agent should measure rather than assume.

\section{Discussion}
\label{sec:discussion}

\subsection{The Client as a First-Order Factor}
Of the three factors RQ1 asks about, only the client is cleanly isolated, and it is the one we find most consequential: for a fixed model, the client driving it had a first-order effect on the solve rate, a 2.1$\times$ gap that the fixes narrowed to roughly 1.5$\times$ but did not close. The model and its client are otherwise confounded, since Claude ran only through Claude Code, so the Claude-versus-DeepSeek range is a combined model-and-client effect. The tool-access regime is the smallest of the three factors, with 5ire the exception. With the model and prompts held constant, the gap is attributable to the client: how it structures the agent loop, parses and repairs tool calls, manages context and retries, and enforces (or fails to enforce) the tool-access constraint. Printer Shares 2 (General, Hard) makes this concrete. After the fixes, all three configurations had the same new \texttt{smb\_ipp\_exploit} tool and the same model. Yet the outcomes diverged: Claude solved it under all three regimes (three calls sufficed in one, 279\,s); 5ire diagnosed the queues and recovered the public half of the secret but stalled at the auth boundary for a partial; and RooCode invoked the tool repeatedly without converging. The tool was necessary but not sufficient, and the client's orchestration decided the outcome. We cannot decompose the gap with one model on two clients, so we state the general claim, that the client matters as much as the model, as a hypothesis rather than a result. It carries a concrete methodological implication: a model's solve rate is meaningful only when the client that produced it is named, and a benchmark that varies the model while fixing an unnamed harness may be measuring the harness as much as the model~\cite{harness-bench,harness-disclosure}.

\subsection{Beyond picoCTF: Decoupling from CTF Conventions}
The orchestrator we evaluated was shaped around CTF conventions, most visibly a hardcoded picoCTF flag format baked into success detection. As part of the intervention we replaced this with a single configurable secret-detection layer. It keeps the picoCTF prefix by default, so the evaluation was unaffected, but it also recognizes real-world indicators such as cloud and API keys, tokens, and private-key headers. The capability tools and fixes were also verified to run against real, non-CTF targets. We are deliberate about what this establishes: the improvements are not CTF-specific at the component level, but this is not an end-to-end measurement of solve rate on real engagements~\cite{bench-practices}, which the flag-bound picoCTF setting cannot provide.

\section{Limitations, Reproducibility, and Ethics}
\label{sec:threats}

\subsection{Threats to Validity}
\textit{Construct.} CTF challenges are a proxy for real offensive-security/DFIR work, not the work itself~\cite{bench-practices}; the flag/secret target and sanctioned environment differ from a live engagement. \textit{Internal.} Fixes were designed and tested on the same challenge set; re-runs cover only previously-unsuccessful trials (one-directional by design: confirms gains, does not test for regressions among passing trials); single-run verdicts outside the stability sample carry the measured run-to-run noise. The intervention is also a bundle, pairing the tool fixes and new tools with a substantial prompt upgrade (the baseline prompt was deliberately minimal), so prompt and tooling contributions cannot be fully separated. The per-trial attribution (79 recoveries without a specialized tool versus 49 with one) bounds how much is attributable to tooling. The inferential tests treat per-configuration trials as independent, whereas the three regimes of a challenge are correlated, so the effective sample size is below the nominal $n = 258$. The effects are large enough (all $p \ll 0.001$) that this does not change the conclusions. The orchestrator we built on also had no test, lint, or type-checking harness, so real defects sat latent: a tool left dead by a shadowed registration, and a server binding every interface against its own loopback default. We fixed both. A tool server's reliability rests as much on a consistent tool-definition contract and a test harness as on catalog breadth. \textit{External.} One benchmark (picoCTF), and a client-as-much-as-model claim supported for one model only and stated as a hypothesis.

\subsection{Reproducibility and Availability}
Results and the fixed fork are available in a public repository~\cite{ourfork}; we do not ship a fully self-contained reproduction package. Upstream HexStrike-AI is MIT-licensed (Muhammad Osama, 0x4m4); our fork retains the notice and we cite the original author~\cite{hexstrike}. For provenance: the base orchestration, wrapped tools, and MCP architecture are HexStrike's; the evaluation design, the fixes and capability tools, and the logging and measurement layer are ours. The environment was Claude (Sonnet 4.6) via Claude Code and DeepSeek \texttt{deepseek-chat} (V3 series, a rolling alias whose snapshot may have shifted across the multi-week study) via RooCode 3.51.1 (VS Code 1.111.0) and 5ire 0.13.2. Sampling used each client's default, with temperature and top-p unpinned. picoCTF instances live 15--30 minutes and the tool/command timeout was 300\,s. Exact trajectory reproduction is thus not guaranteed; the stability sub-study quantifies the variance.

\subsection{Ethics}
All testing was conducted on picoCTF, a sanctioned educational platform, within its authorized-use terms. No live or third-party systems were targeted, and no released artifact contains picoCTF flags.

\section{Conclusion}
Using HexStrike-AI as a testbed, we studied what governs the capability of an LLM security-tool orchestrator and how far targeted engineering recovers its failures, across 774 trials over three model/client configurations and three tool-access regimes.

Relative to prior evaluations that largely report a single solve rate per model~\cite{nyuctf,cybench}, we add three things: the driving client as a first-order factor for a fixed model~\cite{harness-bench,harness-disclosure}, a before-and-after intervention with per-tool attribution of recoveries, and a failure taxonomy whose tool-bound versus reasoning-bound split echoes concurrent penetration-testing analyses~\cite{pentestgpt-v2}.

Three directions follow from this study: an end-to-end evaluation against authorized real-world targets, broadening the client comparison beyond a single model, and an ablation that isolates the prompt upgrade from the new tooling.

\begin{credits}
\subsubsection{\ackname}
This work was supported by the China Scholarship Council (CSC) under Grant No.~2025GSP008568 (R.~Gerard) and Grant No.~2025GSP008567 (A.~Zeghaider).

\subsubsection{\discintname}
The authors have no competing interests to declare that are relevant to the content of this article.
\end{credits}

\bibliographystyle{splncs04}
\bibliography{references}

\end{document}